\documentclass[aps,prd,preprint,showpacs,groupedaddress]{revtex4}

\usepackage{graphicx}

\newcommand{\beq}{\begin{equation}}
\newcommand{\eeq}{\end{equation}}
\newcommand{\bea}{\begin{eqnarray}}
\newcommand{\eea}{\end{eqnarray}}

\def\PRD#1#2#3{Phys. Rev. {\bf D#1}, #2 (#3)}
\def\NPB#1#2#3{Nucl. Phys. {\bf B#1}, #2 (#3)}
\def\PTP#1#2#3{Prog. Theor. Phys. {\bf #1}, #2 (#3)}

\def\EPJC#1#2#3{Eur. Phys. J. {\bf C#1}, #2 (#3)}
\def\PLB#1#2#3{Phys. Lett. {\bf B#1}, #2 (#3)}
\def\PRL#1#2#3{Phys. Rev. Lett. {\bf #1}, #2 (#3)}

\def\r2{\sqrt 2}
\def\PL{\left(\frac{1-\gamma_5}{2}\right)}
\def\PR{\left(\frac{1+\gamma_5}{2}\right)}
\def\la{l_\alpha}
\def\lb{l_\beta}
\def\na{\nu_\alpha}

\def\w{\omega}
\def\x{\chi}
\def\sl{\tilde l}
\def\sn{\tilde \nu }
\def\m#1{{\tilde m}_#1}

\begin{document}

\preprint{
OCHA-PP-250
}

\title{
Generation Mixing of Sneutrinos in Heavier Chargino Decay 
}


\author{
Midori Obara
}
\email[]{midori@mail.ihep.ac.cn}
\affiliation{
Institute of High Energy Physics, 
Chinese Academy of Sciences, 
P.O. Box 918, Beijing 100049, China
}
\author{
Noriyuki Oshimo 
}
\affiliation{
Department of Physics, 
Ochanomizu University, 
Tokyo, 112-8610, Japan
}


\date{\today}

\begin{abstract}

     The heavier chargino decay could yield two charged leptons of 
different generations, owing to generation mixing of sneutrinos.  
We discuss the possibility of producing $e$ and $\mu$ through  
this process in near future collider experiments.  
The analyses are made systematically in the supersymmetric extension of 
the standard model without assuming a specific scenario for the mixing.
Production of the heavier chargino is evaluated in $e^+e^-$ collisions.   
In the parameter region consistent with nonobservation of the radiative 
$\mu$ decay, sizable parts lead to a detectable branching ratio for the 
generation-changing decay of the heavier chargino.

\end{abstract}

\pacs{11.30.Hv, 12.15.Ff, 12.60.Jv, 14.80.Ly}

\maketitle


\section{Introduction}

     The mixing of generations for quarks or leptons is one of 
the topics which have been studied actively in particle physics.   
Examining various phenomena experimentally, generation-changing 
interactions for the quarks are now understood rather well.   
On the other hand, there were before few phenomena 
available for investigating the lepton generation mixing.  
However, neutrino oscillations, which were controversial until recently, 
are now established~\cite{neutrino}, providing information on 
generation-changing interactions for the leptons.   
We have now data on the mixings of both quarks and 
leptons, which can help us to understand underlying theories 
predicting the generation mixings.   

     The generation mixing may also be observed for 
squarks and sleptons whose existence is predicted by the 
supersymmetric extension of the standard model.  
Through loop contributions such mixings could indirectly affect 
various processes, which have 
been already studied extensively~\cite{susyfcnc}.  
For instance, it is known that the masses of the squarks 
or the sleptons with the same weak isospin must be almost degenerate,  
unless the mixing is suppressed.   
If the squarks and sleptons are detected in near future experiments, 
it will become possible to measure their generation-changing interactions 
directly.    
Then, another important clue to theories for the generation mixings can 
be obtained.  
Possible such a process worth being examined is the production of a pair 
of charged leptons belonging to different generations in 
$e^+e^-$ collisions, which is induced by the 
slepton generation mixing.  
In particular, a pair of $\tau$ and $\mu$ or a pair of $\tau$ and 
$e$ could be produced at detectable rates \cite{arkani,porod}.   
  
     In this paper, we study the 
generation-changing process producing a pair of $e$ and $\mu$ 
by the heavier chargino decay, which is due to the sneutrino generation 
mixing.   
Without assuming a specific scenario for the generation mixing, 
we explore parameter space searching for the region which leads 
to a detectable decay rate.   
In order to estimate the heavier chargino production, the cross section 
is evaluated for the process in 
which a pair of different charginos are created in $e^+e^-$ annihilation.  
Differently from the other charged lepton pairs, the possibility of detecting 
$e$ and $\mu$ has not been discussed much in the literature, since 
nonobservation of the radiative decay $\mu\to e\gamma$ may imply a 
small production rate.  
Furthermore, the previous analyses were made under certain scenarios 
for the generation mixings, except for the work by Porod and 
Majerotto~\cite{porod}.   
By model-independent analyses, however, it has been shown recently 
in the pair production of the sneutrinos at $e^+e^-$ colliders 
\cite{oshimo} that the sneutrino 
decays into charged leptons and lighter charginos could yield  
$e$ and $\mu$ at a detectable rate.   
The same generation-changing interactions may also induce  
the heavier chargino decay into a lighter chargino and a pair of 
$e$ and $\mu$, which is mediated by the sneutrinos.  
If the sneutrinos are lighter than the heavier chargino and 
heavier than the lighter chargino, this decay process is 
composed of two successive two-body decays.   
The decay rate could be nonnegligible.  
It will be shown that the branching ratio has a detectable value 
in sizable regions of the parameter space consistent with 
the radiative $\mu$ decay.   
Since the heavier chargino is expected to be copiously produced 
at $e^+e^-$ colliders or hadron colliders, 
the decay will provide a phenomenon for studying 
generation-changing interactions.   

     This paper is organized as follows. In Sect. \ref{sect2}, 
the interactions relevant to our discussions are summarized.      
In Sect. \ref{sect3}, we calculate the generation-changing decay 
width as well as the total width of the chargino.  
In Sect. \ref{sect4}, we make numerical analyses for the 
branching ratio of the heavier chargino decay.   
The cross section of the chargino production is also computed.  
Some discussions are made in Sect. \ref{sect5}.
In Appendices \ref{apexa} and \ref{apexb}, we give the formulae for 
the width of the 
radiative charged lepton decay and the cross section of the 
chargino pair production in $e^+e^-$ annihilation, respectively.  

\section{Generation mixing and interactions \label{sect2}}

     The interactions of charged leptons, sneutrinos, and 
charginos do not conserve generically the generation number.  
This nonconservation arises from  
the generation mixings in the mass matrix $M_l$ for 
the charged leptons $\la$ and the mass-squared 
matrix $\tilde M_\nu^2$ for the sneutrinos $\sn_a$, with $\alpha$ 
and $a$ being generation indices.    
We assume that there is no superfield for the right-handed neutrino 
at the electroweak energy scale.  
These matrices are diagonalized to give the mass eigenstates as  
\bea
U_{lR}^{\dagger} M_l U_{lL}  &=&  {\rm diag} 
\left(m_{l_1},m_{l_2},m_{l_3}\right), 
\\
\tilde{U}_{\nu}^{\dagger} \tilde{M}_{\nu}^2 \tilde{U}_{\nu}  &=&  {\rm diag} 
\left(M_{\tilde{\nu}_1}^2,M_{\tilde{\nu}_2}^2,M_{\tilde{\nu}_3}^2 \right), 
\label{snumass}
\eea
where $U_{lL}$, $U_{lR}$, and $\tilde{U}_{\nu}$ stand for $3\times3$ unitary 
matrices.   
The masses of the charged leptons $e$, $\mu$, and $\tau$ are 
respectively expressed as $m_{l_1}$, $m_{l_2}$, and $m_{l_3}$.   
For the parameters which describe $M_l$ and $\tilde M_\nu^2$, 
we take the mass eigenvalues and the unitary matrices for the 
matrix diagonalization.  
The mass matrix for the charginos $\w_i$ $(i=1,2)$ is given by
\bea
   M^- &=& \left(
      \matrix{           \m2     & -gv_1/\r2 \cr
               -gv_2/\r2 & m_H                   }
           \right).
\label{cmass}
\eea
The $SU(2)$ gaugino mass and the Higgsino mass parameter are denoted by 
$\m2$ and $m_H$, respectively. 
The vacuum expectation values of the Higgs bosons with the 
hypercharges $-1/2$ and $1/2$ are respectively represented by   
$v_1$ and $v_2$, with  $\tan\beta=v_2/v_1$.
The unitary matrices which diagonalize the chargino 
mass matrix are expressed by $C_R$ and $C_L$, 
\bea
C_R^{\dagger} M^- C_L  &=& {\rm diag} \left(m_{\w_1},m_{\w_2} \right),   
\eea
taking $\w_2$ for the heavier chargino. 
The interaction Lagrangian for the mass eigenstates 
$\la$, $\sn_a$, and $\w_i$ is given by
\bea
\cal L &=& i\frac{g}{\r2}\left(V_C\right)_{a\alpha}\sn_a^\dagger\overline{\w_i}
\left[\r2 C_{R1i}^*\PL+C_{L2i}^*\frac{m_{\la}}{\cos\beta M_W}\PR\right]
\la  +{\rm H.c.},
\label{cint}
\eea
with $V_C=\tilde{U}_{\nu}^{\dagger} U_{lL}$. 
The generation mixing is described by the $3\times3$ unitary 
matrix $V_C$, which has four physical parameters.  
For its parameterization, we adopt the same form as the 
standard one for the Cabibbo-Kobayashi-Maskawa matrix, 
\bea
V_C = \left(\matrix{c_{12}c_{13} & s_{12}c_{13} &
                    s_{13}{\rm e}^{-i\delta}  \cr
      -s_{12}c_{23}-c_{12}s_{23}s_{13}{\rm e}^{i\delta} &
      c_{12}c_{23}-s_{12}s_{23}s_{13}{\rm e}^{i\delta} &
          s_{23}c_{13}             \cr
      s_{12}s_{23}-c_{12}c_{23}s_{13}{\rm e}^{i\delta} &
      -c_{12}s_{23}-s_{12}c_{23}s_{13}{\rm e}^{i\delta} &
          c_{23}c_{13}       }    \right),
\eea
with $c_{ij} = \cos \theta_{ij}$ and $s_{ij} = \sin \theta_{ij}$.   
The angles $\theta_{12}$, $\theta_{23}$, and $\theta_{13}$ 
can be put in the first quadrant without loss of generality.    

     The heavier chargino could decay into a charged lepton and a sneutrino  
by the interaction in Eq. (\ref{cint}).  
Another possible mode is the decay into a neutrino and a charged slepton.   
The neutrinos $\na$ are assumed to have Majorana masses.  
Their mass matrix $M_\nu$ is diagonalized by a unitary matrix $U_\nu$,  
\bea
 U_\nu^T M_\nu U_\nu &=& {\rm diag} 
 \left(m_{\nu_1}, m_{\nu_2}, m_{\nu_3}\right).  
\eea
The charged sleptons consist of three left-handed components 
and three right-handed ones.   
The charginos couple to the left-handed components by gauge interactions.    
On the other hand, the couplings of the charginos to the right-handed 
components and the mixings of the left-handed components and the right-handed 
ones are both proportional to the masses of charged leptons.  
If the small effects by the charged lepton masses are neglected, 
we have only to take into consideration the left-handed components.   
Then, the mass-squared matrix $\tilde{M}_l^2$ for the left-handed charged 
sleptons $\sl_a$ is diagonalized by the unitary matrix $\tilde{U}_\nu$ in 
Eq. (\ref{snumass}),    
\bea
\tilde{U}_{\nu}^{\dagger} \tilde{M}_l^2 \tilde{U}_{\nu}  &=&  {\rm diag} 
\left(M_{\sl_1}^2,M_{\sl_2}^2,M_{\sl_3}^2 \right), \\ 
M_{\sl_a}^2 &=& M_{\sn_a}^2-\cos 2\beta M_W^2.  
\eea
The interaction Lagrangian for the mass eigenstates $\na$, $\sl_a$, 
and $\w_i$ is given by
\begin{eqnarray}
\cal L &=& ig\left(V_N\right)_{a\alpha}C_{L1i}
\sl_a^\dagger\overline{\w_i^c}\PL \na  +{\rm H.c.},
\label{eint}
\end{eqnarray}
with $V_N=\tilde U_\nu^\dagger U_\nu$.
The $3\times 3$ unitary matrix $V_N$ describes the generation mixing.
The number of its physical parameters is six, though a definite
parameterization is not necessary for our analyses.

     The heavier chargino could also decay into 
a neutralino and a $W$ boson, or a lighter chargino and a $Z$ boson.  
The mass matrix for the neutralinos $\x_n$ ($n=1$--4) is given by  
\bea
   M^0 &=& \left(
\matrix{     \m1  &               0  & g'v_1/2 & -g'v_2/2 \cr
               0  &              \m2 & -gv_1/2 &  gv_2/2  \cr
 g'v_1/2 & -gv_1/2 &               0  &         -m_H      \cr
-g'v_2/2 &  gv_2/2 &             -m_H &           0         }
           \right),
\label{nmass}
\eea
where $\m1$ represents the $U(1)$ gaugino mass.  
We assume the relation $\m1=(5/3)\tan^2\theta_W\m2$ which is 
suggested by the $SU(5)$ grand unified theory.
The unitary matrix for diagonalizing $M_0$ is expressed by $N$, 
\bea
N^TM^0N &=& {\rm diag} \left(m_{\x_1}, m_{\x_2}, m_{\x_3}, m_{\x_4}\right). 
\eea
The interaction Lagrangian for the mass eigenstates $\w_i$, 
$\x_n$, and $W$ is given by  
\bea
\cal{L} &=& \frac{g}{\r2}\overline{\x_n}\gamma^\mu
\left[G_{Lni}\PL + G_{Rni}\PR \right] \w_i W_\mu^+  +{\rm H.c.},  
\label{wint}\\
 & & G_{Lni} = \r2 N_{2n}^* C_{L1i}+N_{3n}^* C_{L2i}, \quad  
 G_{Rni} = \r2 N_{2n} C_{R1i}-N_{4n} C_{R2i}.
\nonumber 
\eea
The interaction Lagrangian for the mass eigenstates $\w_1$, 
$\w_2$, and $Z$ is given by  
\bea
\cal{L} &=&  \sqrt{g^2+g'^2} \overline{\w_1} \gamma^{\mu} 
\left[ F_L\PL + F_R\PR \right] \w_2 Z_{\mu} + {\rm H.c.},
\label{zint}  \\
 & & F_L = -\frac{1}{2}C_{L21}^* C_{L22}, \quad  
 F_R = -\frac{1}{2}C_{R21}^* C_{R22}.   
\nonumber 
\eea
We do not take into consideration the two-body decay modes for 
the heavier chargino yielding a squark or a Higgs boson, 
assuming that they are not allowed kinematically.  

     The sneutrinos could decay into a charged lepton and a lighter 
chargino by the interaction in Eq.~(\ref{cint}).  
Another possible mode is the decay into a neutrino and a neutralino.   
The interaction Lagrangian for the mass eigenstates $\na$, $\sn_a$, 
and $\x_n$ is given by
\begin{eqnarray}
\cal L &=& i\frac{g}{\r2}\left(V_N\right)_{a\alpha}
\left(-\tan\theta_W N_{1n}+N_{2n}\right)
\sn_a^\dagger\overline{\x_n}\PL \na  +{\rm H.c.},  
\label{nint}
\end{eqnarray}
where $V_N$ has been defined for Eq. (\ref{eint}).

\section{Decay width \label{sect3}}

\begin{figure}
\includegraphics{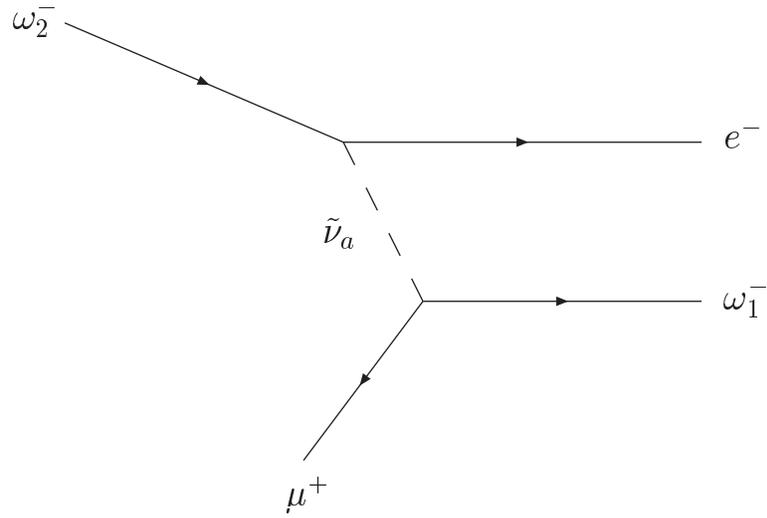}%
\caption{
The Feynman diagram for the heavier chargino decay into  
an electron, a muon, and a lighter chargino.   
\label{diagram1}
   }
\end{figure}

     The heavier chargino decay could yield two charged leptons belonging  
to different generations and a lighter chargino by the   
sneutrino exchanging diagram as shown in Fig. \ref{diagram1}.  
We assume that the three sneutrino masses are almost degenerate 
and the chargino masses satisfy 
the inequality $m_{\w_2} > M_{\sn_a} > m_{\w_1}$. 
Then, the heavier chargino $\w_2$ can decay into a charged lepton $\la$ 
and a sneutrino $\sn_a$, and the sneutrino can decay into a charged 
lepton $\lb$ and a lighter chargino $\w_1$. 
These decay widths are given by
\bea
 \Gamma(\w_2^- \to \la^- \sn_a^*) &=&  |(V_C)_{a\alpha}|^2 
\tilde{\Gamma} (\w_2 \to \la \sn_a), 
\label{widw2}  \\
\Gamma(\sn_a\to \lb^-\w^+_1) &=& |(V_C)_{a\beta}|^2
                           \tilde\Gamma(\sn_a\to \lb\w_1),
\label{widsn}  \\
 \tilde{\Gamma} (\w_2 \to \la \sn_a) &=&  
\frac{g^2}{32 \pi} m_{\w_2} 
\sqrt{\lambda\left(1,\frac{m_{\la}^2}{m_{\w_2}^2},
\frac{M_{\sn_a}^2}{m_{\w_2}^2}\right) } 
\nonumber \\
& &  \biggl[
\left(|C_{R12}|^2+|C_{L22}|^2 \frac{m_{\la}^2}{2\cos^2\beta M_W^2}\right) 
\left(1+\frac{m_{\la}^2}{m_{\w_2}^2}-\frac{M_{\sn_a}^2}{m_{\w_2}^2}\right) 
\nonumber \\
& &  +{\rm Re} [C_{R12} C_{L22}^*] 
\frac{2 \sqrt{2}  m_{\la}^2}{\cos \beta M_W m_{\w_2}} \biggr], 
\nonumber \\
\tilde{\Gamma} (\sn_a \to \lb \w_1)  &=& 
\frac{g^2}{16 \pi} M_{\sn_a} 
\sqrt{\lambda\left(1,\frac{m_{\lb}^2}{M_{\sn_a}^2},
\frac{m_{\w_1}^2}{M_{\sn_a}^2}\right) } 
\nonumber \\
& &  \biggl[
\left(|C_{R11}|^2+|C_{L21}|^2 \frac{m_{\lb}^2}{2\cos^2\beta M_W^2}\right) 
\left(1-\frac{m_{\lb}^2}{M_{\sn_a}^2}-\frac{m_{\w_1}^2}{M_{\sn_a}^2}\right) 
\nonumber \\
&&  -{\rm Re} [C_{R11} C_{L21}^*] 
\frac{2 \sqrt{2}  m_{\lb}^2 m_{\w_1}}{\cos \beta M_W M_{\sn_a}^2}\biggr],
\nonumber 
\eea
where the function $\lambda$ is defined as 
\bea
\lambda(a,b,c) &=& a^2+b^2+c^2-2ab-2bc-2ca.
\eea
The generation-changing decay $\w_2^-\to\la^-\lb^+\w_1^-$ 
$(\alpha\neq\beta)$ is mediated by all the sneutrinos on mass-shell. 
Its width is obtained as  
\bea
& & \Gamma(\w_2^- \to \la^- \lb^+ \w_1^-) =  \sum_{a,b} 
\frac{(V_C^*)_{a\alpha} (V_C)_{a\beta} (V_C)_{b\alpha} (V_C^*)_{b\beta}}
{M_{\sn_a}\Gamma_{\sn_a}+M_{\sn_b}\Gamma_{\sn_b}+ 
i(M_{\sn_a}^2-M_{\sn_b}^2) } 
\nonumber \\ 
& &  \left[ M_{\sn_a} \tilde{\Gamma} (\w_2 \to \la \sn_a) 
\tilde{\Gamma} (\sn_a \to \lb \w_1) 
 +M_{\sn_b} \tilde{\Gamma} (\w_2 \to \la \sn_b) 
\tilde{\Gamma} (\sn_b \to \lb \w_1) 
\right],
\label{width} 
\eea
where $\Gamma_{\sn_a}$ denotes the total decay width of the sneutrino 
$\sn_a$, and the summation for each sneutrino index is done over 
three generations. 
The relations $M_{\sn_a} \gg \Gamma_{\sn_a}$ and 
$|M_{\sn_a}^2-M_{\sn_b}^2| \ll M_{\sn_a}^2 \approx M_{\sn_b}^2$  
have been taken into account.  
The total width of the sneutrino is approximately
determined by the two-body decays, 
\bea
  \Gamma_{\sn_a} &=& \sum_{\alpha}\Gamma(\sn_a\to \la^-\w^+_1) +
             \sum_{\alpha,n}\Gamma(\sn_a\to \na\x_n),  
\eea
where the width of $\sn_a\to \la^-\w_1^+$ is written in Eq. (\ref{widsn}) 
and the width of $\sn_a\to \na\x_n$ is given by 
\bea
      \sum_{\alpha}\Gamma(\sn_a\to \na\x_n) &=& 
           \frac{g^2}{32\pi}M_{\sn_a}\left|-\tan\theta_W N_{1n}+N_{2n}\right|^2
             \left(1-\frac{m_{\x_n}^2}{M_{\sn_a}^2}\right)^2.
\eea
Owing to negligible magnitudes for the neutrino masses, 
the mixing matrix $V_N$ needs not to be specified for obtaining 
the total width.   

     The branching ratio of the generation-changing decay 
depends on the total width of the heavier chargino.   
We assume that the squarks and the Higgs bosons are sufficiently 
heavy not to be produced by two-body decays of the heavier chargino.   
Then, possible two-body modes are the decays into  
$\la\sn_a$, $\na\sl_a$, $\x_nW$, and $\w_1Z$.
The total width of the chargino $\w_2$ becomes approximately 
the sum of these widths, 
\bea
 \Gamma_{\w_2} &=&
 \sum_{\alpha,a} \Gamma(\w_2^- \to \la^- \sn_a^*) + 
 \sum_{\alpha,a} \Gamma(\w_2^- \to \na^* \sl_a^-)  
\nonumber \\
 &+& \sum_{n}\Gamma(\w_2^- \to \x_n W^-) + \Gamma(\w_2^- \to \w_1^- Z),
\eea 
where the width of $\w_2^-\to \la^-\sn_a^*$ is written in Eq. (\ref{widw2}) 
and the widths of other decay modes are given by 
\bea
 \sum_{\alpha} \Gamma(\w_2^- \to \na^* \sl_a^-) &=&
\frac{g^2}{32\pi} m_{\w_2}|C_{L12}|^2
\left(1-\frac{M_{\sl_a}^2}{m_{\w_2}^2}\right)^2,  
\\
 \Gamma(\w_2^- \to \x_n W^-)  &=& \frac{g^2}{64 \pi} m_{\w_2} 
\sqrt{\lambda\left(1,\frac{m_{\x_n}^2}{m_{\w_2}^2},
\frac{M_W^2}{m_{\w_2}^2}\right) } 
\nonumber \\
& &  \biggl[ \left( |G_{Ln2}|^2+|G_{Rn2}|^2 \right) 
\left\{ 1+\frac{m_{\x_n}^2}{m_{\w_2}^2}-2\frac{M_W^2}{m_{\w_2}^2} 
+\left(\frac{m_{\w_2}}{M_W}-  
\frac{m_{\x_n}^2}{m_{\w_2} M_W}\right)^2 \right\} 
\nonumber \\
& & -12{\rm Re}\left(G_{Ln2}G_{Rn2}^*\right)\frac{m_{\x_n}}{m_{\w_2}}
\biggr],   
\\
 \Gamma(\w_2^- \to \w_1^- Z) &=& \frac{g^2+g'^2}{32 \pi} m_{\w_2} 
\sqrt{\lambda\left(1,\frac{m_{\w_1}^2}{m_{\w_2}^2},
\frac{M_Z^2}{m_{\w_2}^2}\right) }  
\nonumber \\
& &  \biggl[ \left( |F_L|^2+|F_R|^2 \right) 
\left\{ 1+\frac{m_{\w_1}^2}{m_{\w_2}^2}-2\frac{M_Z^2}{m_{\w_2}^2} 
+\left(\frac{m_{\w_2}}{M_Z}- 
\frac{m_{\w_1}^2}{m_{\w_2} M_Z}\right)^2 \right\} 
\nonumber \\
& & -12{\rm Re}\left(F_L F_R^*\right)\frac{m_{\w_1}}{m_{\w_2}}
\biggr].     
\eea
The coefficients $G_{Ln2}$, $G_{Rn2}$, $F_L$, and $F_R$ are 
defined in Eqs. (\ref{wint}) and (\ref{zint}).  

     The charged lepton can decay into a lighter charged lepton 
and a photon by exchanging sneutrinos and charginos at the one-loop 
level by the interactions in Eq.~(\ref{cint}).   
The width is obtained by applying the general formula of the 
radiative decay width \cite{bsg} to the relevant interactions.  
This result is written in Appendix \ref{apexa}, which is 
consistent with the formula in the literature \cite{bi}.
The radiative charged lepton decay is also induced by the 
exchanges of charged sleptons and neutralinos.  
However, this neutralino-loop contribution is affected by additional 
factors arising from the right-handed charged sleptons, 
which are not related to the chargino-loop contribution.  
The correlation between the two types of contribution is uncertain.  
Furthermore, the neutralino-loop contribution to the radiative decay 
is generally smaller than the chargino-loop contribution \cite{bsg}.  
It could well happen that the neutralino-loop contribution is negligible.  
Therefore, the interactions in Eq. (\ref{cint}) may be constrained from 
the radiative charged lepton decays by comparing the chargino-loop  
contributions with the experimental bounds.

\section{Numerical analyses \label{sect4}}

     We discuss numerically the decay of the heavier chargino 
$\w_2^-\to e^-\mu^+\w_1^-$.  
The values of the model parameters which prescribe the branching ratio 
of this decay mode 
are constrained by negative results of experimental searches for 
supersymmetric particles and radiative charged lepton decays~\cite{pdg}.    
For the parameters $\m2$, $m_H$, and $\tan\beta$, we take 
four sets of values listed in Table \ref{table1},
where the resultant mass eigenvalues of the charginos and neutralinos 
are also given.  
The mixing matrix $V_C$ and the mass differences between the sneutrinos 
are crucial parameters.  
For simplicity, the mixing angles are taken as the same 
$\theta_{12}=\theta_{23}=\theta_{13}$ $(\equiv\theta)$, with 
the CP violating phase being put at $\delta=\pi/4$, while 
the masses of the sneutrinos $\sn_1$ and $\sn_3$ are fixed at 
$M_{\sn_1}=200$ GeV and $M_{\sn_3}=198$ GeV.  

\begin{figure}
\includegraphics{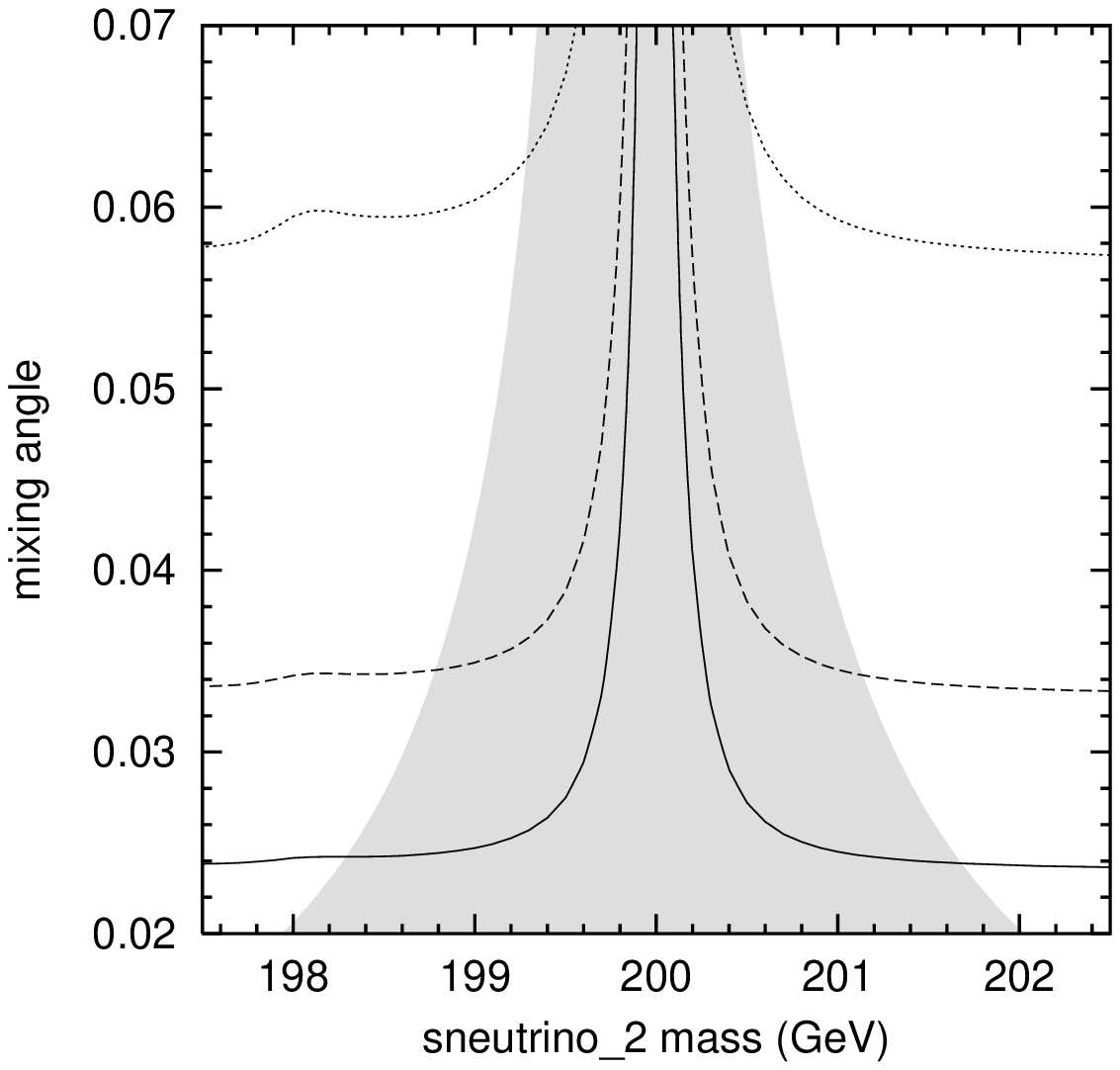}%
\caption{
Contours of the branching ratio BR($\w_2^-\to e^-\mu^+\w_1^-$)  
for the parameter set $(a)$. 
The solid line: BR=$5.0\times10^{-5}$, 
the dashed line: BR=$1.0\times10^{-4}$, 
the dotted line: BR=$3.0\times10^{-4}$. 
The region consistent with the radiative $\mu$ decay is shaded light.   
\label{figa}
   }
\end{figure}

     In Fig. \ref{figa}, we show contours of the branching ratio in the 
$M_{\sn_2}$-$\theta$ plane for the parameter set $(a)$ in Table \ref{table1}.  
The solid, dashed, and dotted lines respectively represent the branching 
ratios $5.0\times10^{-5}$, $1.0\times10^{-4}$, and $3.0\times10^{-4}$.  
Unshaded regions are not allowed by the constraint from 
the radiative decay $\mu\to e\gamma$.    
The other radiative charged lepton decays only impose less tight 
constraints.  
The branching ratio increases, as the mass difference between 
$\sn_1$ and $\sn_2$ or the mixing angle $\theta$ becomes large. 
Similar dependencies also hold for the radiative $\mu$ decay.  
Within the region allowed by the radiative $\mu$ decay, not a small 
part leads to a branching ratio larger than $1.0\times10^{-4}$.  
The mass difference between $\sn_3$ and $\sn_1$ or $\sn_2$ does not 
affect much the branching ratio nor the allowed region.  
Dependence on the mixing matrix is primarily determined by the 
mixing angle $\theta_{12}$.   
A different value for $\theta_{23}$ or $\theta_{13}$ does not alter much 
the resultant branching ratio, as long as these mixing angles are small.  
The results are almost insensitive to the value of the CP violating 
phase $\delta$.   

\begin{figure}
\includegraphics{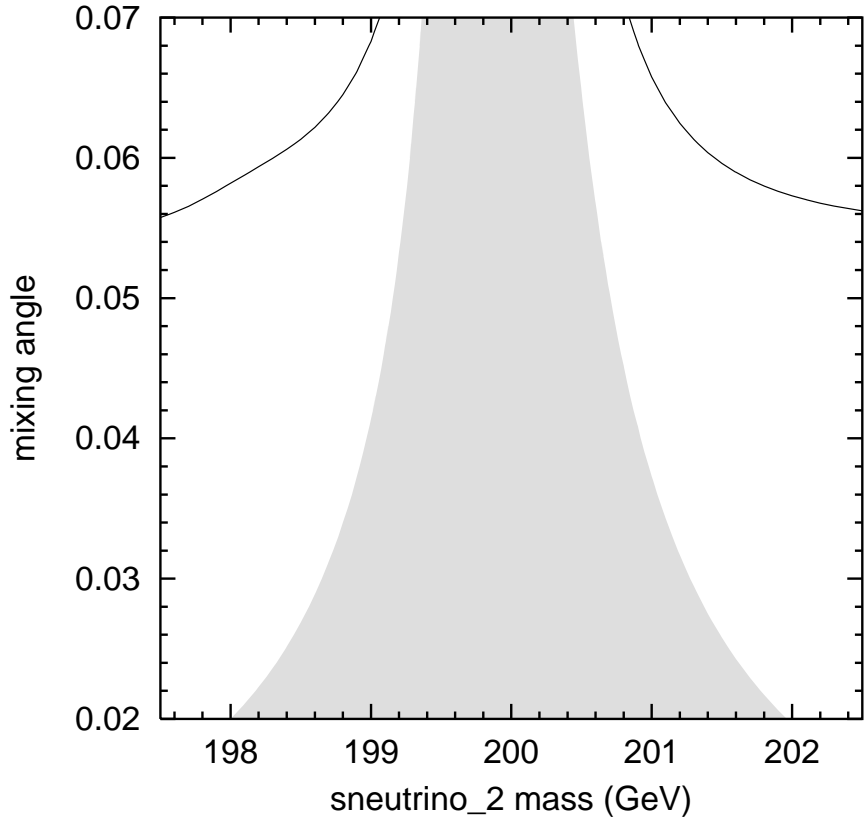}%
\caption{
A contour of the branching ratio BR($\w_2^-\to e^-\mu^+\w_1^-$)  
for the parameter set $(b)$. 
The solid line: BR=$5.0\times10^{-5}$. 
The region consistent with the radiative $\mu$ decay is shaded light.   
\label{figb}
   }
\end{figure}

\begin{figure}
\includegraphics{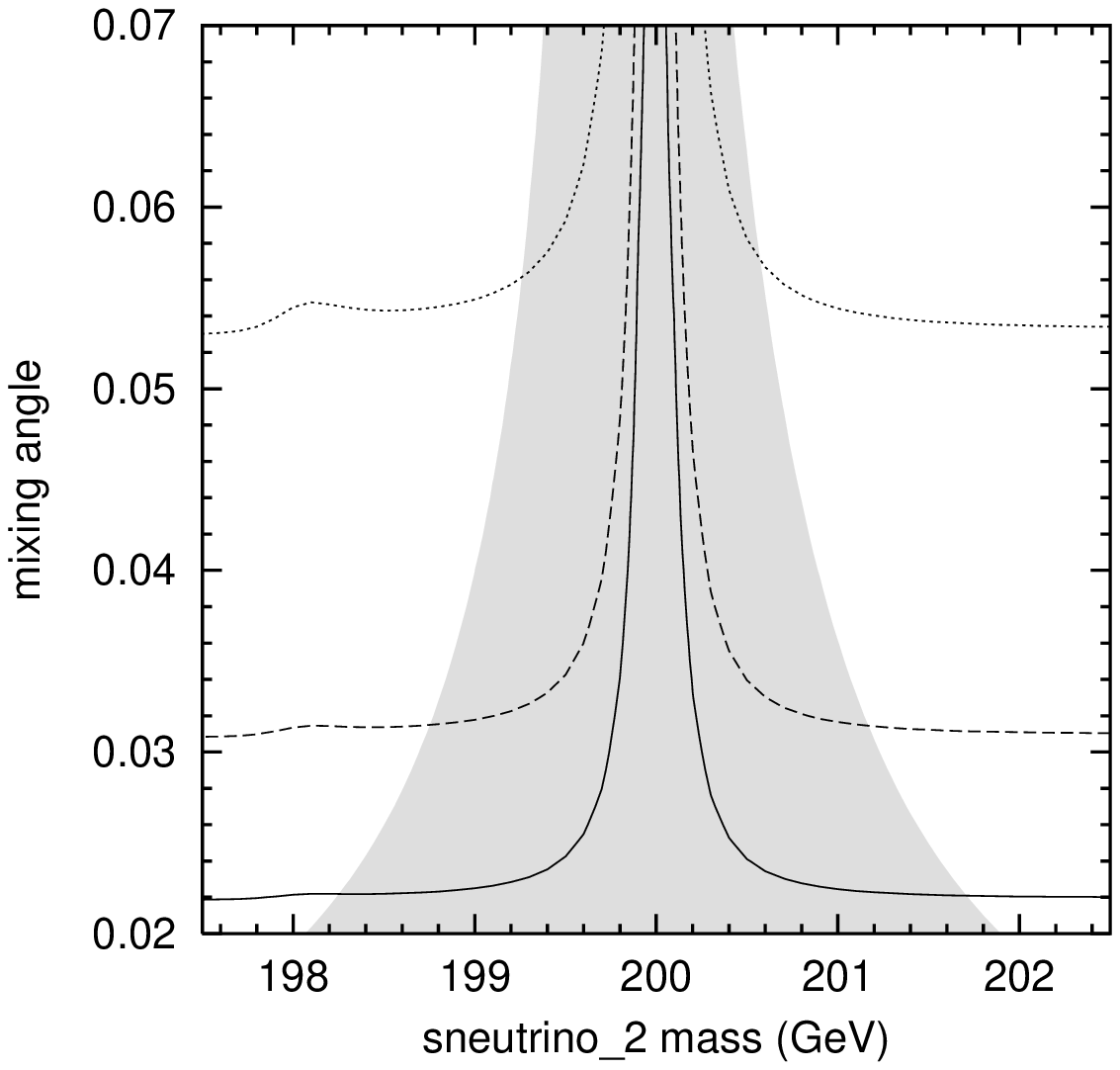}%
\caption{
Contours of the branching ratio BR($\w_2^-\to e^-\mu^+\w_1^-$)  
for the parameter set $(c)$. 
The solid line: BR=$5.0\times10^{-5}$, 
the dashed line: BR=$1.0\times10^{-4}$, 
the dotted line: BR=$3.0\times10^{-4}$. 
The region consistent with the radiative $\mu$ decay is shaded light.   
\label{figc}
   }
\end{figure}

\begin{figure}
\includegraphics{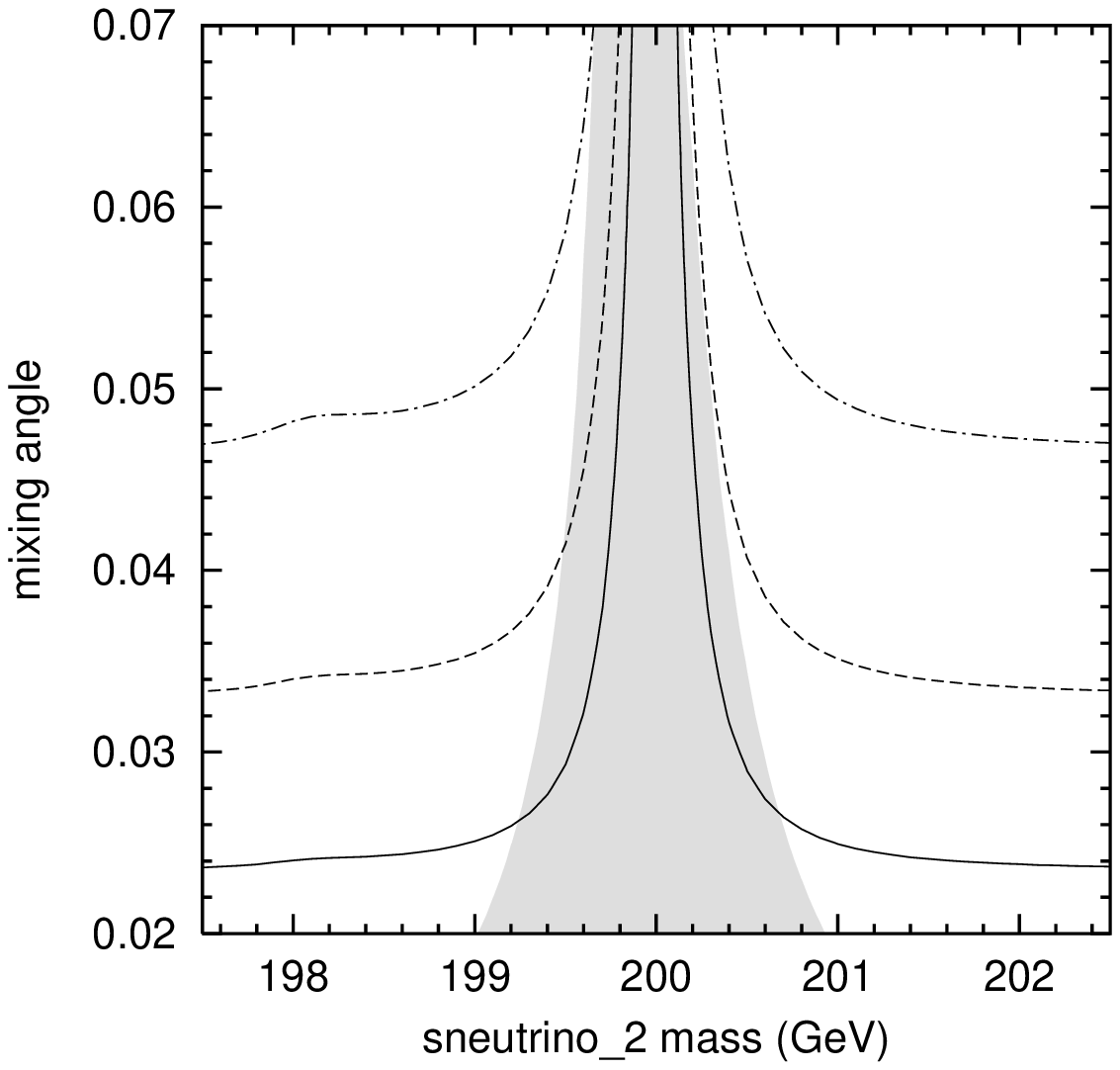}%
\caption{
Contours of the branching ratio BR($\w_2^-\to e^-\mu^+\w_1^-$)  
for the parameter set $(d)$. 
The solid line: BR=$5.0\times10^{-5}$, 
the dashed line: BR=$1.0\times10^{-4}$, 
the dot-dashed line: BR=$2.0\times10^{-4}$. 
The region consistent with the radiative $\mu$ decay is shaded light.   
\label{figd}
   }
\end{figure}

     The branching ratio for the parameter sets $(b)$, $(c)$, and $(d)$ is 
respectively shown in Figs. \ref{figb}, \ref{figc}, and \ref{figd}, 
with the other parameter values being the same as Fig. \ref{figa}.   
The solid, dashed, dot-dashed, and dotted lines represent the branching 
ratios $5.0\times10^{-5}$, $1.0\times10^{-4}$, $2.0\times10^{-4}$,  
and $3.0\times10^{-4}$, respectively.  
Comparing these figures to each other, we can see the dependencies of 
the branching ratio on the parameters $\m2$, $m_H$, and $\tan\beta$.       
Although the region consistent with the radiative $\mu$ decay is not 
much different from each other for a fixed value of $\tan\beta$, 
the branching ratio 
manifestly depends on the relative magnitudes of $\m2$ and $m_H$.  
For a smaller value of $\m2/m_H$, the Higgsino component 
of the heavier chargino is larger.  
Then, the width of the heavier chargino decay into a sneutrino and 
a charged lepton decreases,  
leading to a smaller branching ratio for the generation-changing decay.   
For a larger value of $\tan\beta$, the width of the radiative $\mu$ decay 
is enhanced, since the coupling constants for the right-handed  
charged leptons in Eq. (\ref{cint}) increase.   
Consequently, the allowed parameter region becomes small.  
On the other hand, the branching ratio of $\w_2^-\to e^-\mu^+\w_1^-$ 
does not vary much with the value of $\tan\beta$. 

\begin{figure}
\includegraphics{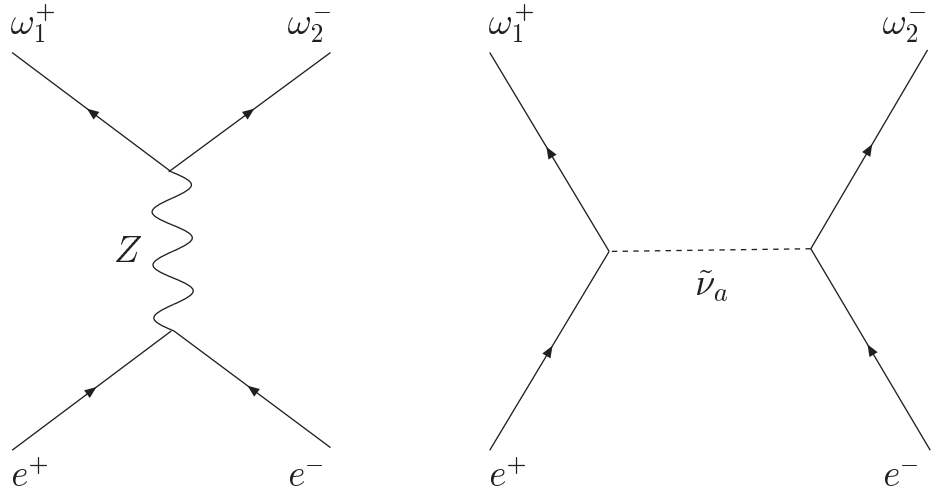}%
\caption{
The Feynman diagrams for the pair creation of a lighter and 
a heavier charginos in $e^+e^-$ annihilation.      
\label{diagram2}
   }
\end{figure}

     We next discuss the cross section of the heavier chargino 
production.  
In $e^+e^-$ collision experiments, the heavier chargino 
is produced as a pair of $\w_1$ and $\w_2$, or a pair of $\w_2^+$ 
and $\w_2^-$, provided that the collider energy is above the threshold.  
In hadron collision experiments, the squark is produced first while  
it decays into a quark and a heavier chargino 
at a large branching ratio.  
In any of these cases, the heavier chargino is expected to be 
produced abundantly.  
For definiteness, we evaluate the cross section 
of the process 
$e^+e^-\to\w_1^+\w_2^-$, which can occur at lower collider energies.   
This production is induced by the diagrams shown in Fig. \ref{diagram2},  
exchanging the $Z$ boson and the sneutrinos.  
The cross section is obtained from the formula for the 
chargino pair production \cite{wino} by converting a pair of the same 
chargino to a pair of $\w_1$ and $\w_2$.   
The result is written in Appendix \ref{apexb}, which 
is consistent with the formula in the literature \cite{choi}.  

\begin{figure}
\includegraphics{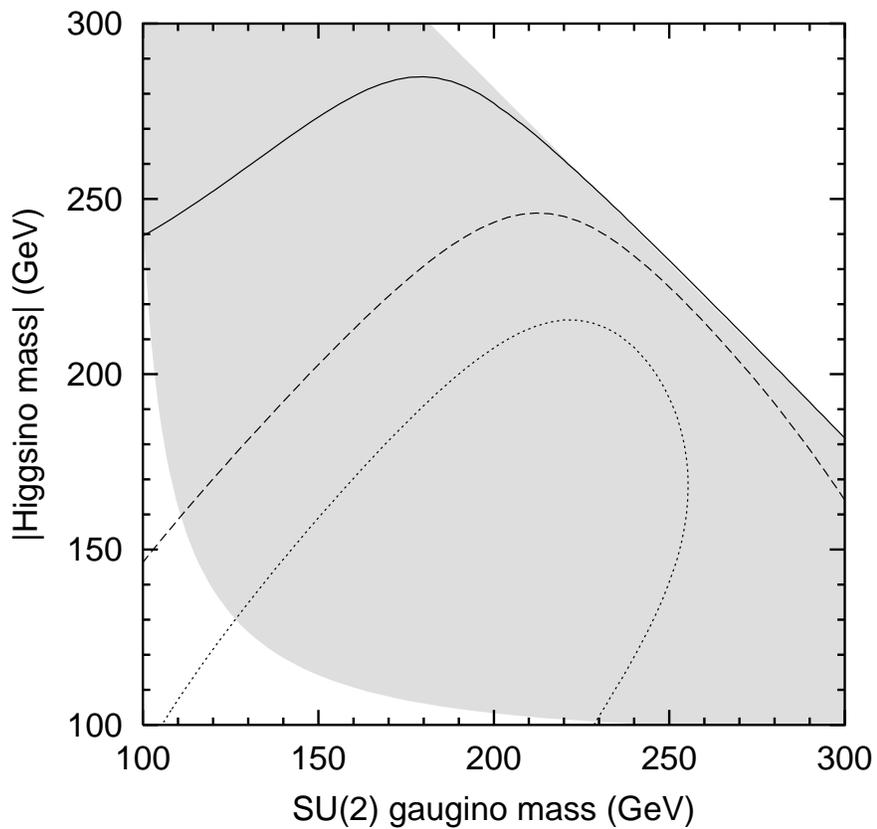}%
\caption{
Contours of the cross section $\sigma(e^+e^-\to\w_1^+w_2^-)$ 
for $\tan\beta=5$ at $\sqrt{s}=500$ GeV.  
The sign of the Higgsino mass is negative.  
The solid line: $\sigma=10$ fb, 
the dashed line: $\sigma=50$ fb, 
the dotted line: $\sigma=100$ fb. 
Outside the light shaded region, the lighter chargino mass is 
smaller than 100 GeV, or the production of a pair of different 
charginos is not allowed kinematically.  
\label{crs}
   }
\end{figure}

     In Fig. \ref{crs}, contours of the cross section are shown 
in the $\m2$-$|m_H|$ plane for $\tan\beta=5$, where the value of 
$m_H$ is negative.  
All sneutrino masses are taken equal to 200 GeV and the 
generation mixing matrix $V_C$ is assumed to be a unit matrix.  
The total energy in the center-of-mass frame is set at $\sqrt{s}=500$ GeV.  
The solid, dashed, and dotted lines respectively represent 
the cross sections 10, 50, and 100 fb.   
In the left-lower unshaded region, the mass of the lighter 
chargino is less than 100 GeV, which is ruled out by 
LEP experiments.  
In the right-upper unshaded region, the sum of the masses for the heavier 
and lighter charginos exceeds 500 GeV, so that the pair 
production is not allowed kinematically.   
In a sizable region of the parameter space, the cross section becomes  
of order of 100 fb.  
In table \ref{table2}, the cross sections for the 
parameter values of Figs. \ref{figa}$-$\ref{figd} are 
given at the total energies 500, 600, and 700 GeV.

\section{Discussions \label{sect5}}

     We have calculated the branching ratio of $\w_2^-\to e^-\mu^+\w_1^-$ 
and the cross section of $e^+e^-\to \w_1^+w_2^-$.  
It should be noted that the heavier chargino also has a decay mode 
$\w_2^-\to \mu^-e^+\w_1^-$ with almost the same branching ratio.  
Furthermore, there exist charge conjugate decays which have 
the same branching ratios.  
The cross section of $e^+e^-\to \w_1^-w_2^+$ is equal to  
that of $e^+e^-\to \w_1^+w_2^-$.  
Since the lighter chargino decays dominantly into two quarks 
and the lightest neutralino, the final state of these successive 
production and decay processes mostly consists of a pair of 
$e$ and $\mu$ and four jets with missing energy-momentum.  
For the integrated luminosity 100 fb$^{-1}$ of $e^+e^-$ 
collisions, a few events or more are expected in sizable regions of 
the model parameter space.   
Although the event number is not large, the signature is distinctive.  
The generation mixing of the sneutrinos could be explored in    
the production and decay of the heavier chargino.  

     There are several processes which eventually lead to the same 
topology as the discussed generation-changing process.  
Suppose that a pair of sleptons are produced in $e^+e^-$ collisions.  
Possible decay modes of the sneutrinos or the charged sleptons 
contain $\sn_a\to\la\w_1$ or $\sl_a\to\la\x_2$, respectively.  
The second lightest neutralino $\x_2$ dominantly decays into 
two quarks and the lightest neutralino, giving the same final topology 
as the lighter chargino.  
Generation-changing interactions of the sleptons could thus yield 
the final state of $e$, $\mu$, and four jets with missing energy-momentum.  
If the energy of $e^+e^-$ collisions is sufficiently higher than 
the primarily produced particle masses, the slepton decays give 
one charged lepton in each hemisphere, whereas two charged leptons are 
contained in one hemisphere for the heavier chargino decay.  
These processes could be distinguished from each other by the angular 
distributions of the final charged leptons.  
On the other hand, if the collision energy is not so high, the primarily 
produced particles are approximately at rest, so that their decays allow 
the charged leptons to go in any direction.  
Then, the distinction between the above processes becomes difficult.    
We could only know that the generation-changing processes are 
generated by the slepton interactions.  

     Serious experimental backgrounds receive contribution 
from generation-conserving interactions through the production 
of $\tau^+$ and $\tau^-$ and their subsequent leptonic decays.   
For instance, after $e^+e^-$ collisions the following reactions 
may occur:
$\w_1^+\w_2^-\to \tau^+\tau^-\w_1^+\w_1^-$,   
$\sn_\tau^*\sn_\tau\to \tau^+\tau^-\w_1^+\w_1^-$, and  
$\sl_\tau^*\sl_\tau\to \tau^+\tau^-\x_2\x_2$ 
by the supersymmetric model interactions and 
$ZW^+W^-\to \tau^+\tau^-W^+W^-$   
by the standard model interactions.     
Their final states could contain $e$, $\mu$, and four jets 
with missing energy-momentum.  
However, in these processes $e$ and $\mu$ are produced by the 
three-body decays of $\tau$.  
In the generation-changing processes, on the other hand, 
$e$ and $\mu$ are produced by the two-body decays.   
Assuming the collision energy is not so high and 
thus $\w_2$ is nonrelativistic,   
the energies of $e$ and $\mu$ are roughly monochromatic.   
The magnitudes and the distributions of the charged lepton energies 
are very different from the generation-conserving processes.  
In particular, if the masses of the charginos and the sneutrinos 
are known, the energies of $e$ and $\mu$ can be specified.   
Appropriate energy cuts would enable reduction of the backgrounds.

\begin{acknowledgments}
     M.O. would like to thank A. Sugamoto and G.C. Cho for encouragement.
The work of N.O. is supported in part by the Grant-in-Aid for
Scientific Research on Priority Areas (No. 16028204) from the
Ministry of Education, Science and Culture, Japan.
\end{acknowledgments}

\appendix

\section{\label{apexa}}

     The width for the radiative charged lepton decays $\mu \to e \gamma$, 
$\tau \to e \gamma$, and $\tau \to \mu \gamma$ is given by  
\bea
\Gamma(\lb \to \la \gamma) = \frac{\alpha_{EM}}{2}m_{\lb}
\Bigl( |F_2 |^2 + |G_2 |^2 \Bigr)
\biggl( 1-\frac{m_{\la}}{m_{\lb}} \biggr)^2
\biggl( 1-\frac{m_{\la}^2}{m_{\lb}^2} \biggl),
\eea
where $F_2$ and $G_2$ represent the dipole form factors, 
\bea
F_2 &=& \frac{g^2}{32 \pi^2}\sum_{a,i} \frac{m_{\la}  +m_{\lb}}{M_{\sn_a}}
(V_C)_{a\beta} (V_C^*)_{a\alpha}
\biggl[ \left( \left|C_{R1i}\right|^2
+ \frac{\left|C_{L2i}\right|^2 m_{\la} m_{\lb}}{2\cos^2\beta M_W^2} \right)
\frac{m_{\la}  +m_{\lb}}{M_{\sn_a}} 
I_1 \left(\frac{m_{\w_i}^2}{M_{\sn_a}^2}\right)
\nonumber \\
&&  + \left( \frac{C_{R1i}^* C_{L2i}  m_{\la}}{\r2\cos\beta M_W}
+ \frac{C_{R1i} C_{L2i}^*  m_{\lb}}{\r2\cos\beta M_W} \right)
 \frac{m_{\w_i}}{M_{\sn_a}} 
I_2 \left(\frac{m_{\w_i}^2}{M_{\sn_a}^2}\right) \biggr]
\\
G_2  &=&  \frac{g^2}{32 \pi^2}\sum_{a,i} \frac{m_{\la} +m_{\lb}}{M_{\sn_a}}
(V_C)_{a\beta} (V_C^*)_{a\alpha}
\biggl[ \left( \left|C_{R1i}\right|^2
- \frac{\left|C_{L2i}\right|^2 m_{\la} m_{\lb}}{2\cos^2\beta M_W^2} \right)
\frac{m_{\lb} -m_{\la}}{M_{\sn_a}} 
I_1 \left(\frac{m_{\w_i}^2}{M_{\sn_a}^2}\right)
\nonumber \\
&&  - \left( \frac{C_{R1i}^* C_{L2i} m_{\la}}{\r2\cos\beta M_W}
- \frac{C_{R1i} C_{L2i}^*  m_{\lb}}{\r2\cos\beta M_W} \right)
\frac{m_{\w_i}}{M_{\sn_a}} 
I_2 \left(\frac{m_{\w_i}^2}{M_{\sn_a}^2}\right) \biggr].
\eea
The functions $I_1$ and $I_2$ are defined as 
\bea
I_1 (r) &=& \frac{1}{12 (1-r)^4} \, (2+3r-6 r^2+r^3+6r \log r), \\
I_2 (r) &=& \frac{1}{2 (1-r)^3} \, (-3+4r-r^2-2 \log r).
\eea

\section{\label{apexb}}

     The production cross section for a pair of light and heavy charginos 
in $e^+ e^-$ collisions is, at the total energy 
$\sqrt{s}$ in the center-of-mass frame, given by  
\bea
 & & \sigma(e^+ e^- \to \w_1^+ \w_2^-) 
= \frac{1}{4\pi s^2} \int_{t_-}^{t_+} dt  
\nonumber \\
& & \Biggl[ \frac{(g^2+g'^2)^2}{4(s-M_Z^2)^2}
\Biggl\{ \left(|A_L|^2 |F_R|^2+|A_R|^2 |F_L|^2\right) 
\left(t-m_{\w_1}^2\right)\left(t-m_{\w_2}^2\right) 
\nonumber \\
& &  +\left(|A_L|^2 |F_L|^2+|A_R|^2 |F_R|^2\right) 
\left(t+s-m_{\w_1}^2\right)\left(t+s-m_{\w_2}^2\right) 
\nonumber \\
& &  +2 \left(|A_L|^2+|A_R|^2\right) 
{\rm Re}[F_L F_R^*] m_{\w_1}m_{\w_2} s \Biggr\} 
\nonumber \\
& &  +\frac{g^4}{16} \left( 
 \sum_{a} \frac{|(V_C)_{a1}|^2}{t-M_{\sn_a}^2} \right)^2 
|C_{R11}|^2 |C_{R12}|^2 \left(t-m_{\w_1}^2\right)\left(t-m_{\w_2}^2\right)
\nonumber \\
& &  -\frac{(g^2+g'^2)g^2}{4(s-M_Z^2)} 
 \sum_{a} \frac{|(V_C)_{a1}|^2}{t-M_{\sn_a}^2}A_L 
\nonumber \\
& & \left\{ {\rm Re}\left[F_R^* C_{R11}^* C_{R12}\right] 
\left(t-m_{\w_1}^2\right)\left(t-m_{\w_2}^2\right) 
+ {\rm Re}\left[F_L^* C_{R11}^* C_{R12}\right]  
m_{\w_1} m_{\w_2} s \right\} \Biggr],
\label{crossw1w2} \\ 
& & A_L = -\frac{1}{2}+\sin^2\theta_W, \quad  A_R = \sin^2\theta_W,   
\nonumber 
\eea
where the coefficients $F_L$ and $F_R$ are defined in Eq. (\ref{zint}).  
The range of the integration is expressed by $t_\pm$ as   
\bea
t_{\pm}  &=&  \frac{1}{2} 
\left[ m_{\w_1}^2+m_{\w_2}^2-s \pm 
\sqrt{\lambda(s,m_{\w_1}^2,m_{\w_2}^2)}\right]. 
\eea
The mass of the electron has been neglected.


\newpage
\pagebreak


%

\begin{table}
\caption{
The parameter values and the resultant masses for the charginos 
and neutralinos.  The unit of mass is GeV.
\label{table1}
 }
\begin{ruledtabular}
\begin{tabular}{ccccc}
  & $(a)$ & $(b)$ & $(c)$ & $(d)$    \\
\hline
 $\tan\beta$ & 5 & 5 & 5 & 10 \\
 $\m2$  &   200   &  150  &   250  &   200  \\
 $m_H$  & $-200$ & $-250$ & $-150$ & $-200$  \\
\hline
 $m_{\w_1}$ & 166 & 144 & 144  & 159  \\
 $m_{\w_2}$ & 256 & 278 & 278  & 260  \\
\hline
 $m_{\x_1}$ &  94 &  72 & 108  &  92  \\
 $m_{\x_2}$ & 164 & 143 & 149  & 158  \\
 $m_{\x_3}$ & 215 & 265 & 166  & 213  \\
 $m_{\x_4}$ & 252 & 271 & 277  & 258  \\
\end{tabular}
\end{ruledtabular}
\end{table}

\begin{table}
\caption{
The cross section of $e^+e^- \to \w_1^+\w_2^-$ at $\sqrt{s}$ 
of different energies (GeV) 
for the parameter values $(a)$--$(d)$, corresponding respectively to 
Figs. \ref{figa}--\ref{figd}.   
The unit of cross section is fb.
\label{table2}
 }
\begin{ruledtabular}
\begin{tabular}{ccccc}
 $\sqrt{s}$ &  $(a)$ & $(b)$ & $(c)$ & $(d)$    \\
\hline
 500 & 109 & 18 & 102 & 112  \\
 600 & 108 & 20 &  97 & 109  \\
 700 &  93 & 18 &  81 &  94  \\
\end{tabular}
\end{ruledtabular}
\end{table}


\newpage
\pagebreak

%

\end{document}